\newcommand{\DocVersion}{To appear in {\it The Astronomical Journal} -- July 2004}

\documentclass[10pt,preprint2]{aastex}

\slugcomment{\DocVersion}

\shortauthors{Shuping, Morris, \& Bally}
\shorttitle{Mid-Infrared mosaic of Orion BN/KL}

\newcommand{\kms}{km s$^{-1}$}

\newcommand{\cmthree}{cm$^{-3}$}

\newcommand{\water}{H$_2$O}
\newcommand{\msun}{M$_{\sun}$}
\newcommand{\lsun}{L$_{\sun}$}

\newcommand{\htwo}{H$_2$}

\newcommand{\vlsr}{$V_{lsr}$}

\newcommand{\nhthree}{NH$_3$}
\newcommand{\thetaonec}{$\theta^1$ C Orionis}
\newcommand{\Lp}{L$^{\prime}$}

\begin{document}

\title{A new mid-infrared map of the BN/KL region using the Keck telescope}

\author{R. Y. Shuping\altaffilmark{1} \& Mark Morris\altaffilmark{2}}
\affil{Div. of Astronomy \& Astrophysics, UCLA \\
MS 8371, Los Angeles, CA 90095-1562  \vspace{10pt}}
\and
\author{John Bally\altaffilmark{3}}
\affil{ Center for Astrophysics \& Space Astronomy, University of Colorado  \\
   Campus Box 389, Boulder, CO 80309-0389}

\altaffiltext{1}{{\tt shuping@astro.ucla.edu}}
\altaffiltext{2}{{\tt morris@astro.ucla.edu}}
\altaffiltext{3}{{\tt John.Bally@colorado.edu}}

\begin{abstract}
\small
We present a new mid-infrared (12.5 $\mu$m) map of the \objectname{BN/KL} high-mass star-forming complex in Orion using the LWS instrument at Keck I.  Despite poor weather we achieved nearly diffraction-limited images (FWHM $= 0.38$\arcsec) over a roughly 25\arcsec$\times$25\arcsec\ region centered on  IRc2 down to a flux limit of $\approx 250$ mJy.  Many of the known infrared (IR) sources in the region break up into smaller sub-components.  We have also detected 6 new mid-IR sources.  Nearly all of the sources are resolved in our mosaic.  The near-IR source ``n'' is slightly elongated in the mid-IR along a NW--SE axis and perfectly bisects the double-peaked radio source ``L''.  Source n has been identified as a candidate for powering the large IR luminosity of the BN/KL region ($L = 10^5$~\lsun).  We postulate that the 12 $\mu$m emission arises in a circumstellar disk surrounding source n.  The morphology of the mid-IR emission and the Orion ``hot core'' (as seen in NH$_3$ emission), along with the location of water and OH masers, is very suggestive of a bipolar cavity centered on source n and aligned with the rotation axis of the hypothetical circumstellar disk.  \objectname{IRc2}, once thought to be the dominant energy source for the BN/KL region, clearly breaks into 4 sub-sources in our mosaic, as seen previously at 3.8 -- 5.0~\micron.  The anti-correlation of mid-IR emission and NH$_3$ emission from the nearby hot core indicates that the IRc2 sources are roughly coincident (or behind) the dense hot core.  The nature of IRc2 is not clear:  neither self-luminous sources (embedded protostars) nor external heating by source I can be definitively ruled out.  We also report the discovery of a new arc-like feature SW of the \objectname{BN} object, and some curious morphology surrounding near-IR source ``t".  
\end{abstract}


\newpage
\section{Introduction}
\label{sect:Intro}

    Because of its proximity \citep[450 pc,][]{Genzel+Stutzki89}, the
      OMC 1 region, which includes the Becklin-Neugebauer object and
      the Kleinmann-Low nebula (BN/KL) and the Orion ``hot core'', is among the most studied
      regions of high-mass star formation in the sky.  
   An outstanding uncertainty about this region is the source of the
       $\sim 10^5$~\lsun\ of infrared (IR) luminosity from BN/KL.
   Though many individual peaks of thermal infrared (7 to 24 $\mu$m)
      emission
        have been identified \citep{Lonsdale+82}, many may not be
      self-luminous.
    The VLA studies by \citet{Menten+Reid95} indicate that, in
      addition to
        the BN object, the region contains at
      least two
        additional ultra-compact radio sources, I and L, separated
        by about 3\arcsec\ from each other (1,500~AU in projection).
Radio source L is coincident with IR source ``n'' \citep{Lonsdale+82}.
   Recent mid-IR studies of the region suggest that the hot stars
      powering these compact \ion{H}{2} regions may be sufficient to
      drive the enormous IR flux from BN/KL \citep{Gezari+98}.

The Orion ``hot core'' is a warm, dense cloud ($T > 220$ K, $n($\htwo$) = 5 \times 10^7$ \cmthree) in the SE region of BN/KL, first observed in ammonia emission \citep{Morris+80}.  An unusually high number of hydrogen-saturated molecules are observed toward the hot core, perhaps due to the evaporation of icy grain mantles \citep{Blake+87,Brown+88}.  
The velocity centroid of the ``shell masers'' masers surrounding radio source I is nearly identical to the LSR velocity of the hot core ($v_{lsr} = 5$\kms), which suggests that the two are physically related.
The {\em absence} of any IR emission and the morphology of the ammonia emission in the vicinity also suggests that radio source I is embedded in the hot core \citep{Gezari+98,Wilson+00}.  Hence source I is probably responsible for most of the heating and chemistry in the hot core. 

  Sources I and n are both associated with strong OH, H$_2$O,
        and SiO maser emission \citep{Johnston+89,Genzel+81}.
An expanding arcminute-scale complex of high velocity ($v$ =
        30 to 100~\kms ) H$_2$O masers surrounds the entire  \objectname{OMC 1}
      region, with a center of expansion that coincides, within several
      arcseconds, with source n \citep{Genzel+81,Menten+Reid95}.
In addition to this high-velocity flow, a low-velocity 
       arcminute-scale cluster of much brighter maser spots is
      associated with this expansion ($v_{expansion} = 18 \pm$2~\kms).
As shown by
        \citet{Gaume+98}, the so-called 22-GHz \water\ ``shell''
        masers, which were mostly resolved-out in the VLBI observations
        of \citet{Genzel+81}, are concentrated in a 2\arcsec\
        by 0.5\arcsec\ strip centered on source I and are oriented roughly
        orthogonal to the bipolar CO outflow emerging from OMC 1
        \citep{Chernin+Wright96}.  
Bright SiO maser emission lies within
        0.1\arcsec\ of source I, and consists of four clusters of maser spots
      having velocities very similar to those of the H$_2$O shell
      masers.
\citet{Greenhill+98} and \citet{Doeleman+99}
        found that these masers are concentrated into four linear
      chains located
        north, east, south, and west of source I.
Contrary to prior conclusions, a new analysis by
      \citet{Greenhill+03} suggests that these masers trace a disk seen
      roughly edge-on with a projected rotation axis at PA $\sim 45$\arcdeg\ and a
      bipolar wind impacting the surrounding dense medium.

Radio CO line emission indicates that there is also a fast (30 to 100~\kms), weakly collimated
      bipolar outflow
        emerging from the OMC 1 region with a blue-shifted lobe toward
      the northwest.
The OMC 1 outflow
        has a mass of about 10~\msun\ and a kinetic energy of about
        $4 \times 10^{47}$ ergs \citep{Kwan+Scoville76}. 
There are also ``fingers'' of \htwo\ emission, presumably from shocked gas, pointing away from the BN/KL region to the NW and SE.
The OMC 1 outflow
        and the \htwo\ fingers \citep{Allen+Burton93,Stolovy+98,Schultz+99,Kaifu+00} appear to indicate that
        a powerful ejection occurred in OMC 1 relatively recently.
The \htwo\ finger system consists of over a hundred individual
      bow shocks
        which delineate a relatively wide-angle bipolar outflow
        toward PA $\approx$ 315\arcdeg\ with an opening angle
        of more than 1 radian in each lobe.
The proper motion vector field of the \htwo\ line-emitting knots, as observed using {\it Hubble
      Space Telescope} (HST), has been interpreted as an explosion that occurred about
        1010$\pm$140 years ago \citep{Lee+Burton00,Doi+02}.  This is consistent with the OMC 1 outflow observed in CO.

In this paper we present the highest spatial resolution map at 12.5
     ~\micron\ of the BN/KL region to date.\footnote{
     Note added during review:  Another dataset obtained with LWS on the Keck I telescope has just been published by \citet{Greenhill+04} and covers some of the same fields in BN/KL as presented here.}
In the next section we describe our observations using the LWS
      instrument on Keck I and the data analysis applied to produce the
      mosaic.
In Section~\ref{sect:Results} we present our results, including 6
      new mid-IR sources, and new morphologies for IRc2 and IR source n.
In Section~\ref{sect:Discussion} we discuss the interpretation of
      some of our results, especially for sources n, I, and IRc2.  We
      also try to rationalize the various outflows in the region and
      some curious features surrounding near-IR source ``t''.
Summary and conclusions are given in
      Section~\ref{sect:Summary+Conclusions}.

\section{Observations \& Data Analysis}
\label{sect:Obs+DataAnalysis}

Observations of the BN/KL region were made using the Keck
          Observatory facility mid-IR camera LWS on UT 16 November 2002.
LWS is a mid-IR imaging and spectroscopy instrument mounted on the
          forward Cassegrain focus of Keck I, employing a Boeing $128
          \times 128$ As:Si BIB array with a $10.2 \times 10.2$\arcsec\ field of view \citep{Jones+Puetter93}. 
Weather conditions were poor on UT 16 November 2002:  the thermal background and atmospheric transmission varied by $\sim 50$\% throughout the first half of the night.  
Approximately 30 individual observations (frames) were made
          of the BN/KL region at slightly overlapping positions using
          the 12.5\micron\ filter (12 -- 13~\micron\ bandpass with $> 80$\% transmission).
The chopping secondary mirror was driven at 2 Hz with a
          30\arcsec\ E--W throw. 
Each frame was observed using the standard mid-IR chop-nod
          technique with two chopping positions ``+'' (on-source) and
          ``-'' (off-source); after chopping with the source in
          chop-beam ''+'', the telescope was nodded along the chop axis
          so that the object would sit in chop-beam "-", and chopping
          would continue. 
Each frame was observed for one complete chop-nod cycle,
          yielding a total on-source integration time of 27.6 seconds
          per mosaic frame.  
The standard stars $\beta$ Peg, $\beta$ And, and $\beta$ Gem
          were also observed throughout the night for PSF determination
          and flux calibration.
Our data are nearly diffraction limited at 12.5~\micron\ with
          PSF FWHM $= 0.38$\arcsec and a Strehl ratio of 35\%.

\begin{figure}
\epsscale{0.9}
\plotone{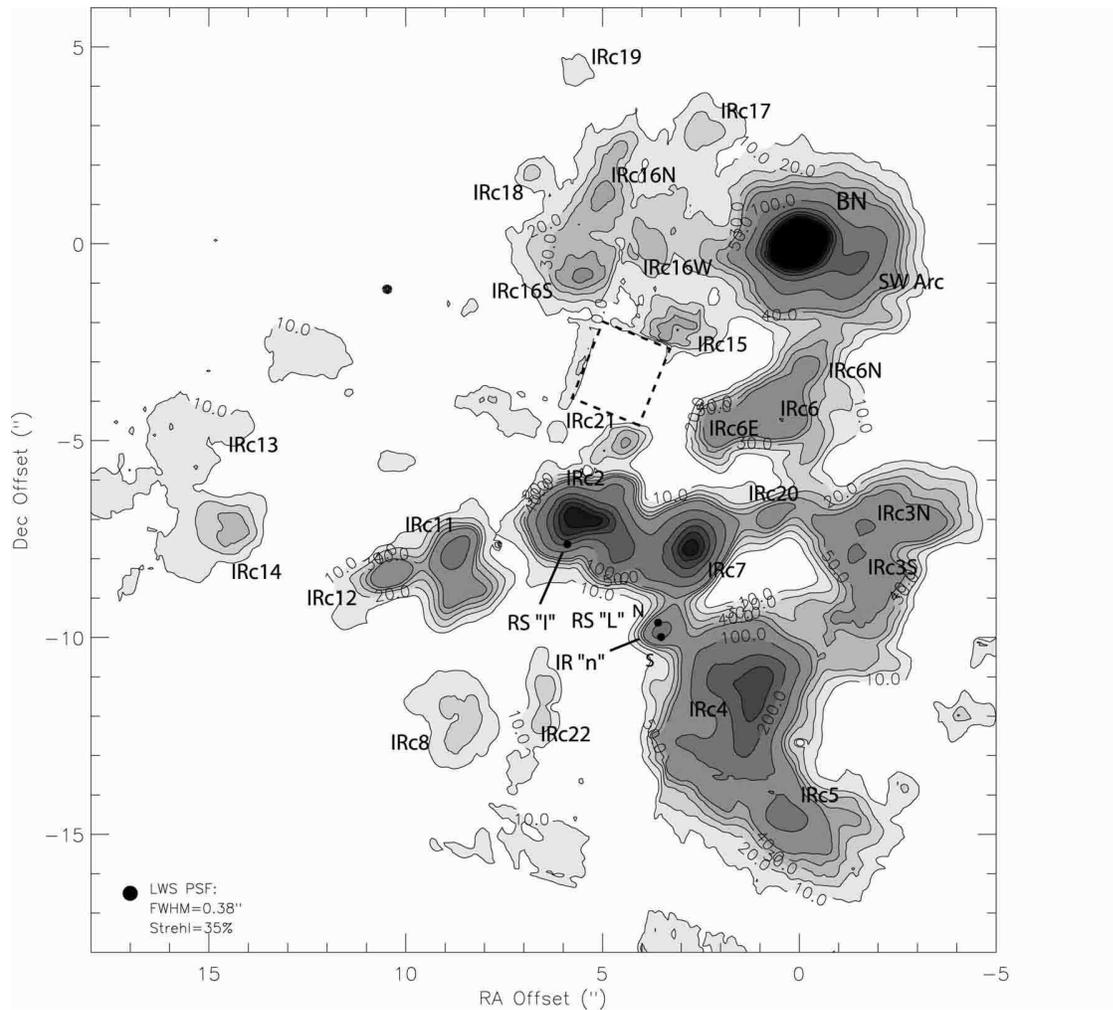}
\caption{Filled contour map of the Keck/LWS 12.5~\micron\ mosaic for BN/KL.  Flux is mJy per $0.081 \times 0.081$\arcsec\ pixel.  Contours shown are for 10, 20, 30, 40, 50, 100, 200, 300, 400, 500, 600, and 6000 mJy; representative contours are marked in the figure.  The dashed box shows the area not covered by our observations.  Coordinates are centered on the BN object (RA = 05h35m14.117s,  DEC = -05\arcdeg22\arcmin22.90\arcsec [J2000]). See Table 1 for IR source identifications.  Continuum radio sources (RS)``I'' and ``L'' are also indicated \citep{Menten+Reid95}.  }
\label{fig:bnkl_irsources}
\end{figure}

The LWS instrument saves all coadded data in individual chop
          and nod beams for each observation frame.
The data were reduced by first finding and eliminating bad
          pixels in the raw data and then differencing the chop beams
          for each nod position.  
There were very few contaminating sources in our off-source chop beams:  in these cases we merely threw out the offending chop pair and used the chop pair from the other nod position.
Typically there was some residual sky signal left over after
          differencing.  In order to create a smooth mosaic, we opted to eliminate this residual by calculating a sky mode and subtracting the value from the differenced images.  While this effectively reduces the background to near zero, it also eliminates any diffuse nebular emission from the region.
A gain map for the detector was created using an average of the off-source
          chop positions for each chop-nod cycle, correcting for the
          dark current which we measured at the start of the night. 
          This gainmap was then divided into the differenced chop-beam
          data.  
We found that, after nodding, the telescope did not always
          return the target to the exact same position on the detector,
          so we registered the nod position data using the
          cross-correlation of the two images and then coadded to
          create a final image for each telescope pointing.  
Reduced images for each pointing were then combined into a
          mosaic by registering on sources common to successive frames.
The final mosaic is shown in Figure~\ref{fig:bnkl_irsources}.
Because of our frame overlapping strategy, we inadvertently
          missed one small region of BN/KL inbetween IRc2 and BN
          (shown as a dashed box in Figure~\ref{fig:bnkl_irsources}).

A linear astrometrical solution for the completed mosaic was
          determined using the coordinates for BN \citep{Menten+Reid95}
          and IRc7 \citep{Gezari+98}.
The plate scale was determined to be 0.0805 \arcsec\ pixel$^{-1}$.
Astrometrical errors were determined from the positional
          errors for BN and IRc7  \citep{Gezari+98}, the centroiding
          accuracy ($< 0.5$ pixel), and the offset from IRc7 or BN
          (whichever is closer).  Positional errors range from 0.1\arcsec\
 near IRc7 and BN to 0.3\arcsec\ at the edge of the
          mosaic.
No attempt has been made to measure or correct for optical
          distortion, which is assumed to be small over the (very
          narrow) field of view.

Due to highly variable atmospheric transmission, flux
          calibration using the observed standards was not reliable.
Instead we calibrated the entire mosaic using known fluxes
          for BN and IRc2 from \citep{Gezari+98}.  
Since the instrumental response varied from frame to frame in
          the mosaic, the flux calibration for any frame that does not
          contain BN or IRc2 can deviate significantly from those
          frames that do.
The regions around IRc2, IRc7, and BN were observed multiple
          times, allowing us to determine a maximum variation
          ($2\sigma$) in the flux solution from frame to frame of 
          $\sim 25$\%.  
The uncertainty in the known fluxes of BN and IRc2 at 12
         ~\micron\ is roughly 10\% \citep{Gezari+98}, so the absolute flux
          calibration error ($\approx 2\sigma$) for our mosaic ranges
          between 10 and $\sim 27$\%.  
The 3$\sigma$ flux limit for our data is approximately 250
          mJy for a compact source measured in a 15-pixel-radius
          aperture.

\section{Results}
\label{sect:Results}

The 12.5~\micron\ mosaic of BN/KL presented here is the
          highest spatial resolution mid-IR map of the region to date.  
Even at this resolution (PSF FWHM $= 0.38$\arcsec) all the
          detected sources are resolved -- there are no point sources
          down to our flux limit.  
Many known IR sources do break up into multiple compact
          sources, but even these sub-components are resolved.  
We have also detected 6 new mid-IR sources, one near BN which
          we have called the ``southwest arc'', and 5 more which we have
          named IRc18 through 22, following \citet{Gezari+98}.
Positions for each of the sources relative to BN are given in
          Table~\ref{table:sources}.  The uncertainty in the relative
          position is dominated by the offset from IRc7 or BN, as
          discussed in Section~\ref{sect:Obs+DataAnalysis}, and ranges from
          0.1\arcsec\ to 0.3\arcsec.  
We have also measured the flux for a handful of the most
          compact sources, using a large circular aperture and an outer
          annulus to determine the background flux density and its variance
          (Table~\ref{table:sources}).  The flux uncertainty is
          dominated by variations caused by poor weather  conditions
          (Section~\ref{sect:Obs+DataAnalysis}) but is not larger than $\approx 
          27$\%.

The mid-IR emission from the BN/KL region is generally
          thought to fall into two categories:  Relatively dense clouds
          heated by external UV radiation from local high-mass stars
          embedded in the region, and dense regions heated by embedded
          YSOs.  
Most of the mid-IR sources (except for BN and n) are not seen
          at wavelengths less than a few microns owing to the significant
          (but patchy) foreground extinction associated with OMC 1.
IRc3, 4, 5, and 6 have relatively cool mid-IR colors and
          polarizations consistent with external illumination
          \citep{Gezari+98, Dougados+93,Minchin+91}, most likely from
          sources I and n.
In our data, IRc3 and 6 break up into smaller fragments, but
          there is no evidence of  point sources which might be
          interpreted as embedded young stars.
BN, source n, and IRc2, 7, 11, and 12 all have compact
          quasi-elliptical morphologies suggestive of dense clumps with
          embedded YSOs.  
In addition, these sources have much warmer mid-IR colors
          indicative of embedded objects (though the clumps would also
          be heated by external UV radiation) \citep{Gezari+98}.

\begin{figure}
\epsscale{1.0}
\plottwo{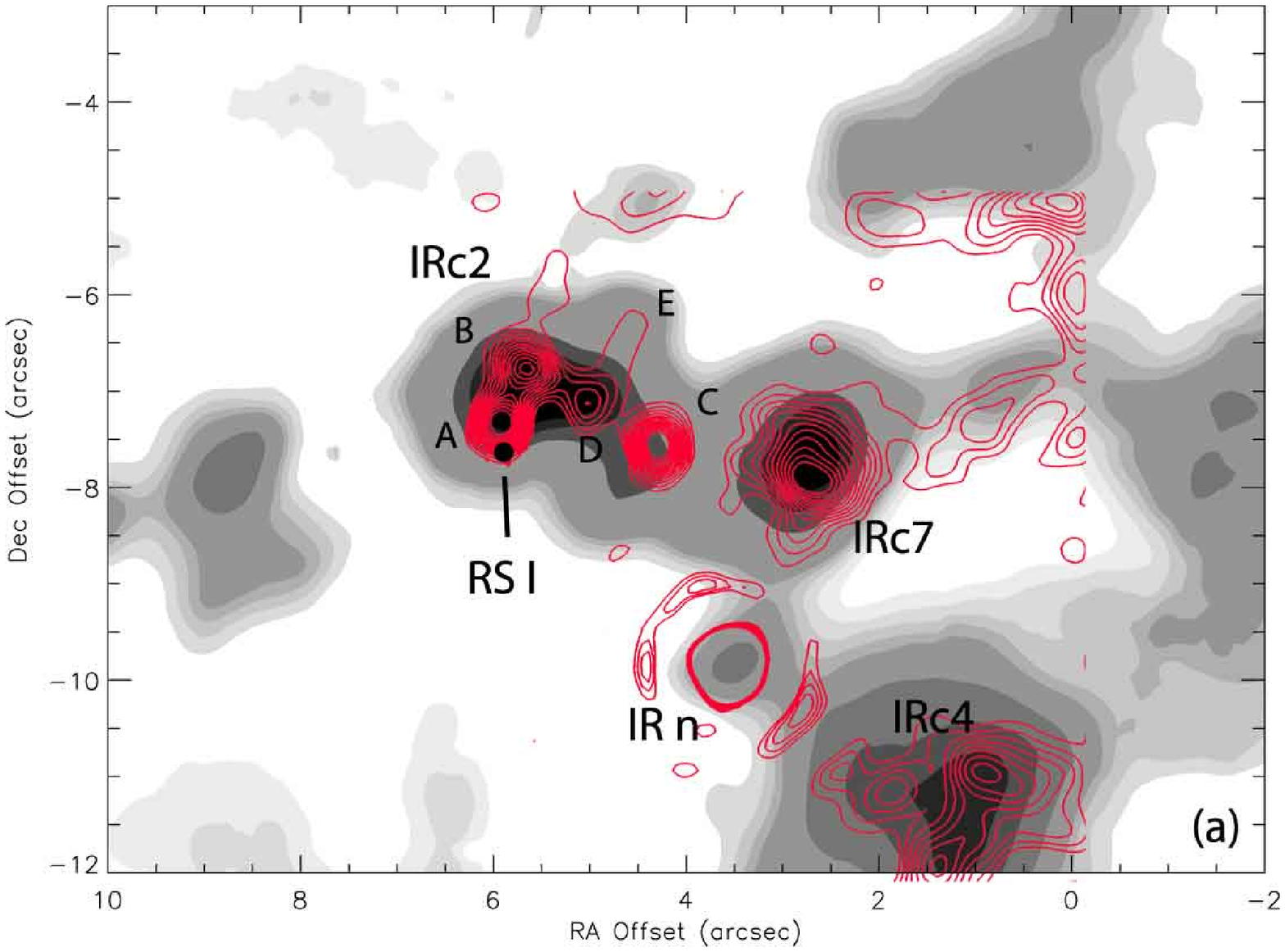}{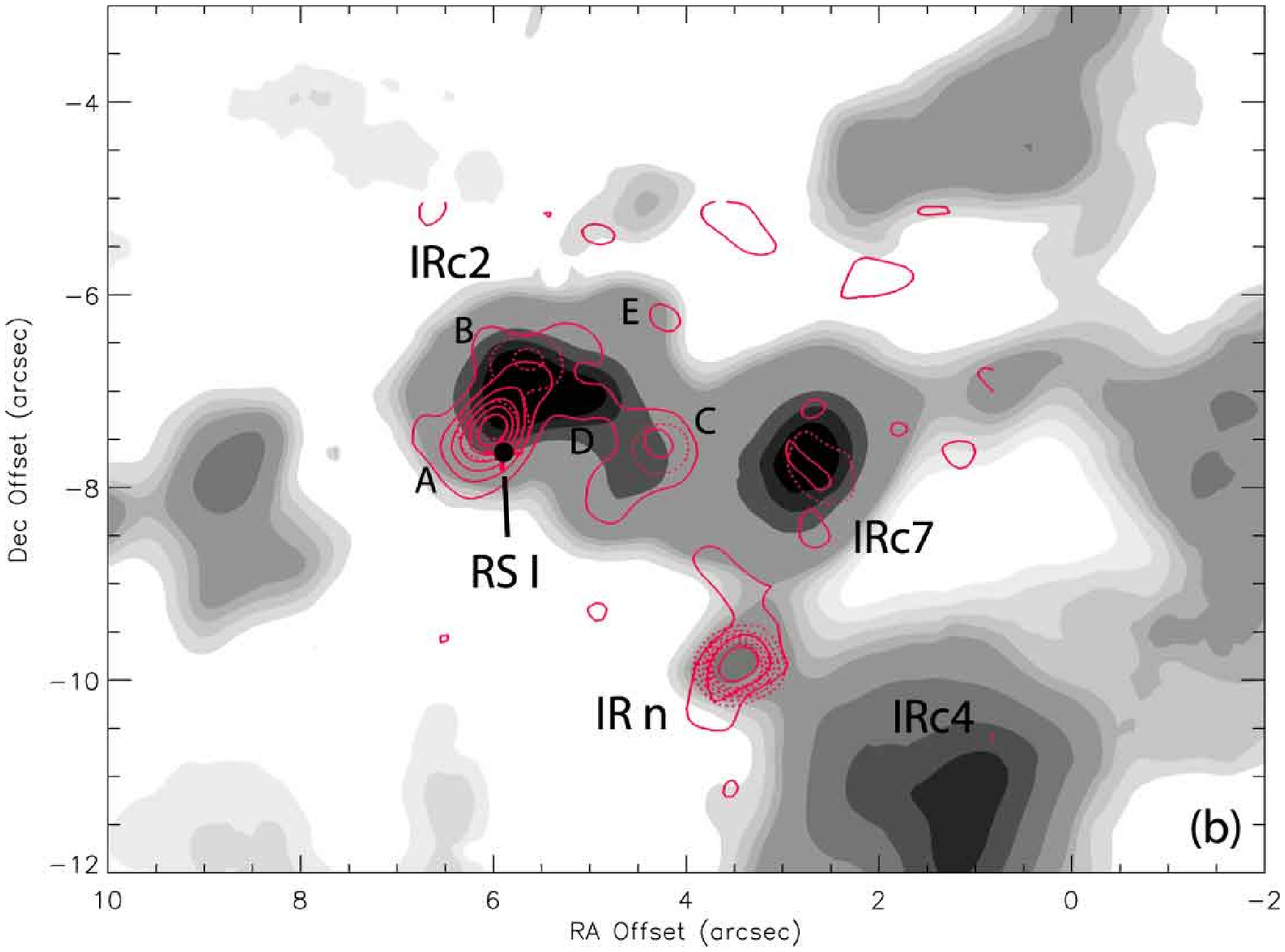}
\caption{Filled contour map of the 12.5\micron\ mosaic in the IRc2 region (grayscale, same contours as Fig.~\ref{fig:bnkl_irsources}) overlaid with (a) \Lp\ emission and (b) 4.0 (dashed) and 5.0 (solid)~\micron\ emission \citep{Dougados+93} in red contours.  The spatial resolution of the $\sim 4$\micron\ data is FWHM $= 0.5$\arcsec. See Table 1 for IR source identifications.   }
\label{fig:IRc2_12-4}
\end{figure}

\begin{figure}
\epsscale{1.0}
\plottwo{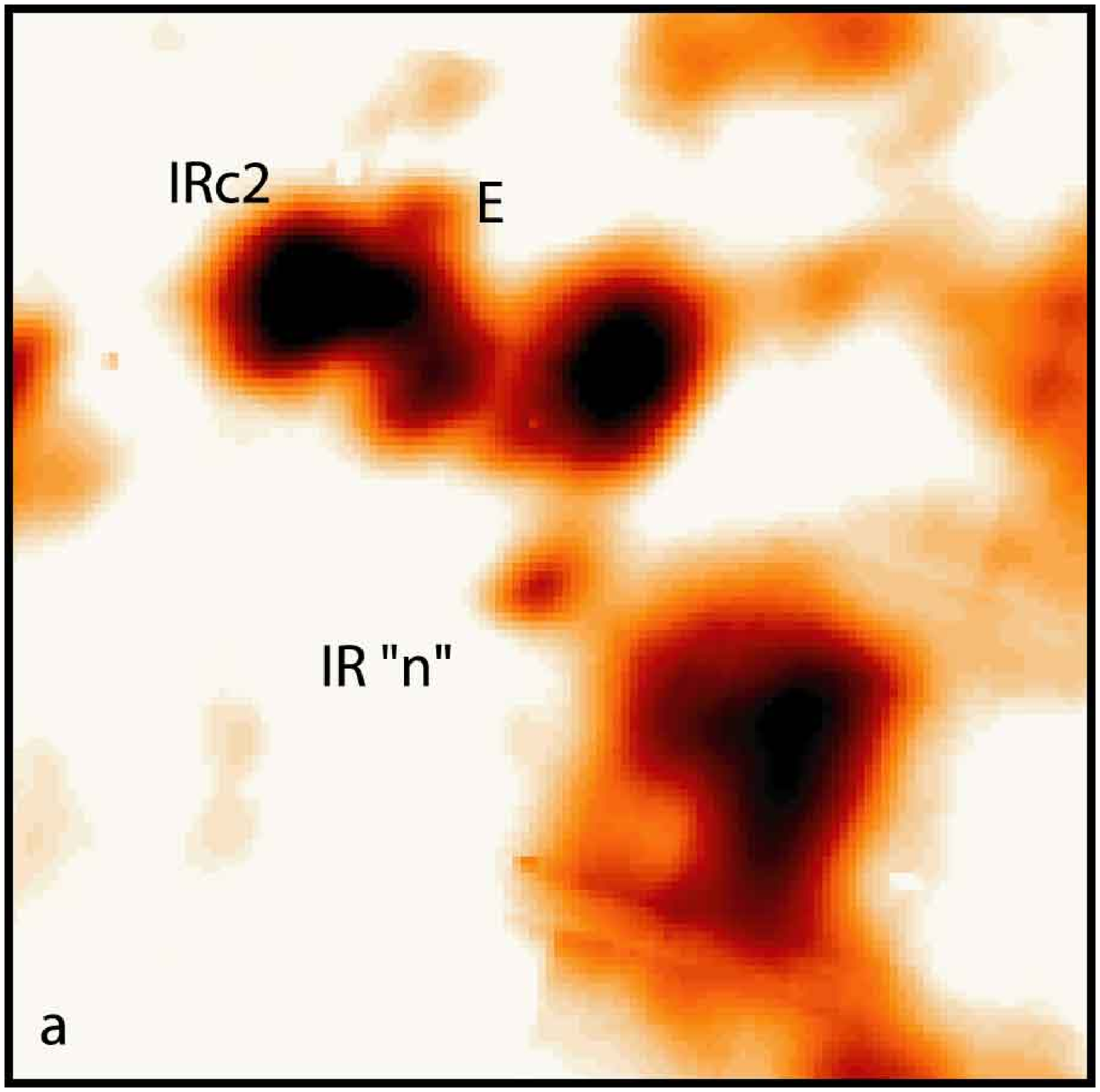}{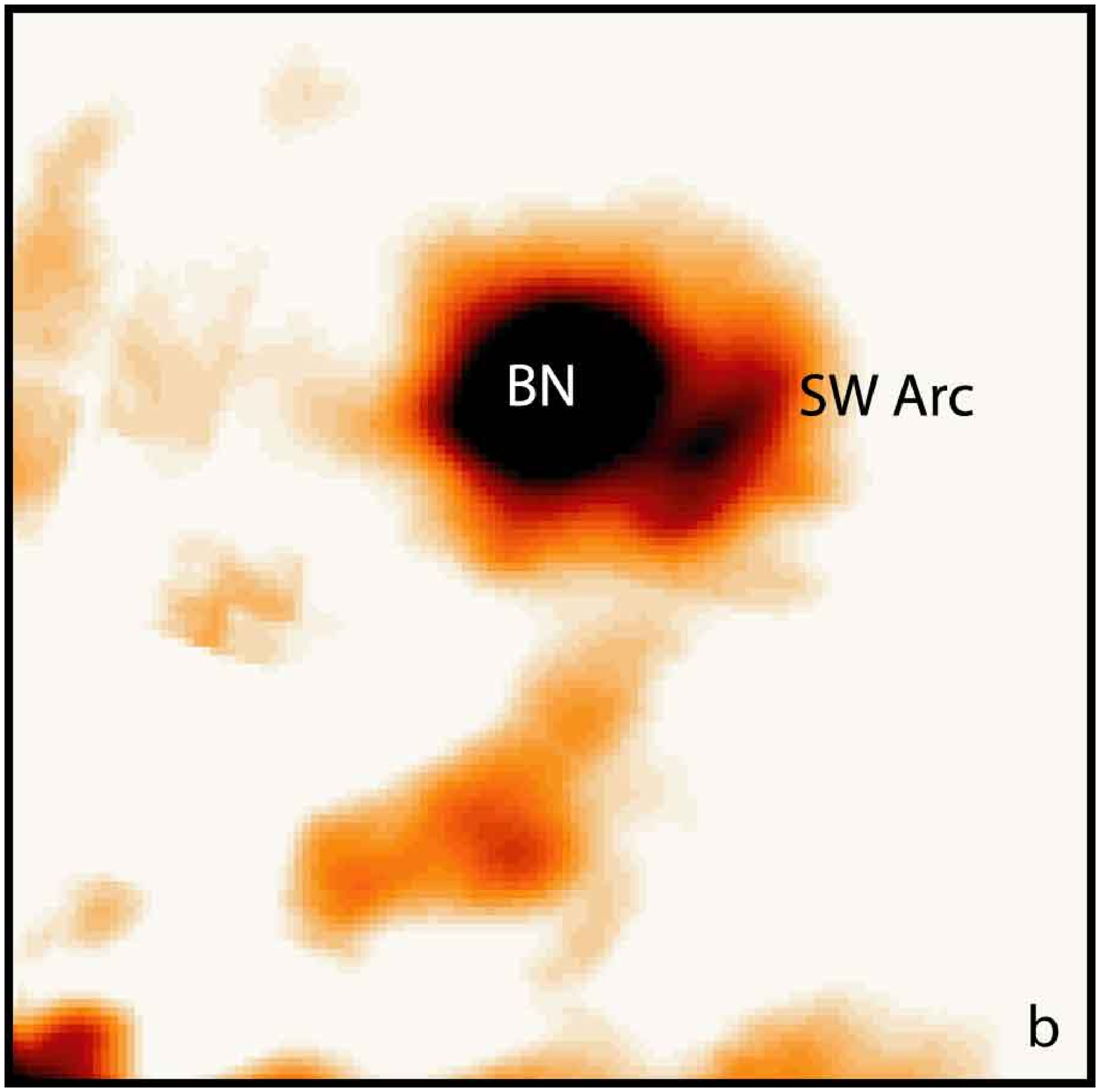}
\caption{Detailed 12.5~\micron\ sub-images (10\arcsec\ $\times$ 10\arcsec) enhancing low level emission around near-IR source n (a) and the BN SW arc (b).}
\label{fig:map_detail}
\end{figure}

The structure of IRc2 at 12~\micron\ is similar to that seen at 4 -- 5~\micron, but not identical (Figure~\ref{fig:IRc2_12-4}).  Four unresolved sources  (A -- D) have been detected at \Lp -- with marginally resolved counterparts at 4.0 and 5.0~\micron \citep{Dougados+93}.  The morphology of IRc2 is clearly wavelength dependent from 3.8 to 12.5~\micron:  the eastern peak of IRc2 at 12.5~\micron\ lies between A and B; and the peak of 12.5~\micron\ emission near source C is offset from that source by about 0.5\arcsec\ to the SE.  
The faint emission just north of source D, first identified as a ``jet-like'' feature by \citet{Dougados+93}, shows up as a distinct compact source at 12.5~\micron\ which we have named ``E'' (Figure~\ref{fig:map_detail}).  Note that none of these IRc2 sources have been detected at 2~\micron\ \citep[e.g.,][]{Lonsdale+82,Schultz+99}. 
Though very close (within 0.5\arcsec), source A is not coincident with source I. 
The nature of IRc2 and radio source I will be discussed in
          Section~\ref{sect:Discussion}.

IR source n coincides with
          the double radio continuum source ``L'' (N and S),
          which has been identified as a possible source of significant
          ionizing radiation in the BN/KL region \citep{Menten+Reid95}.
Source n appears as an unresolved point source throughout the
          near-IR \citep{Lonsdale+82,Dougados+93}.
At 12~\micron, however, source n appears elongated along a
          NW-SE position angle (Figure~\ref{fig:map_detail}), perfectly bisecting the northern and
          southern components of radio source L
          (Figure~\ref{fig:bnkl_irsources}) -- suggestive of a
          circumstellar disk straddled by ionized bipolar cavities (or lobes, see Section~\ref{sect:Discussion}).
The uncertainty in the 12~\micron\ position of source n (0.1\arcsec) is smaller than the $\sim 0.2$\arcsec\ separation
          between n and both components of L.

\begin{figure}
\epsscale{0.75}
\plotone{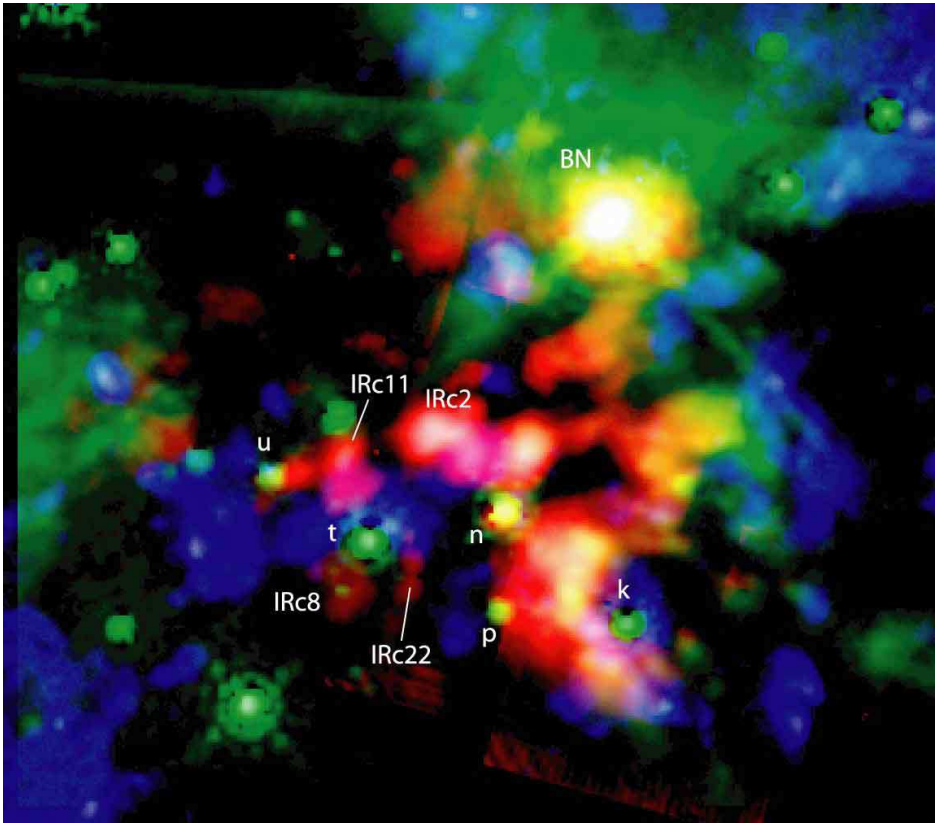}
\caption{Three-color composite image of BNKL region (31\arcsec\ $\times$ 27\arcsec):  Red = 12.5\micron\ (this work); Green = NICMOS 2.15~\micron\ continuum; Blue = 2.12~\micron\ \htwo\ emission (NICMOS N212 $-$ N215).  A logarithmic stretch has been applied to each channel independently.  To avoid confusion, stellar features in the \htwo\ channel (blue) have been masked out.  The NICMOS data were obtained from the Multi-Mission Archive at Space Telescope and were first published by \citet{Schultz+99}.
}
\label{fig:BNKL_12-N215-H2_ann}
\end{figure}

Though limited to one bandpass, the increased sensitivity of
          our dataset over previous efforts has revealed some new
          mid-IR sources in the BN/KL region:  IRc18 -- 22 (following \citet{Gezari+98})  and the ``BN SW Arc''.
The concentric relationship to BN suggests that the SW Arc is a nearby shell of heated gas and dust.  Due to the inhomogeneous nature of the BN/KL region and the lack of data at other wavelengths, we can only speculate on the nature of the SW arc:  it may be externally heated by UV radiation or is possibly a portion
of a shocked, compressed shell created by an outflow or jet from BN.  
Source IRc18 is compact but still resolved, and may have a
          counterpart at 4~\micron \citep{Dougados+93}.
IRc19 is an irregularly shaped source just southwest of IR
          source ``q'' \citep{Lonsdale+82}; it is not clear whether the
          two are related.
Source IRc20 appears as a ``bridge'' between IRc7 and 3. 
          This feature is also seen clearly at 4~\micron\
          \citep{Dougados+93}.
IRc21 is just north of IRc2 and may be associated with a
          bright knot of \htwo\ emission immediately to the west
          (see Figure~\ref{fig:BNKL_12-N215-H2_ann}).
IRc22 appears to be associated with IR source ``t''
          \citep{Lonsdale+82} and will be discussed further below.

\section{Discussion}
\label{sect:Discussion}

\subsection{IR Source n}

\begin{figure}
\epsscale{0.9}
\plotone{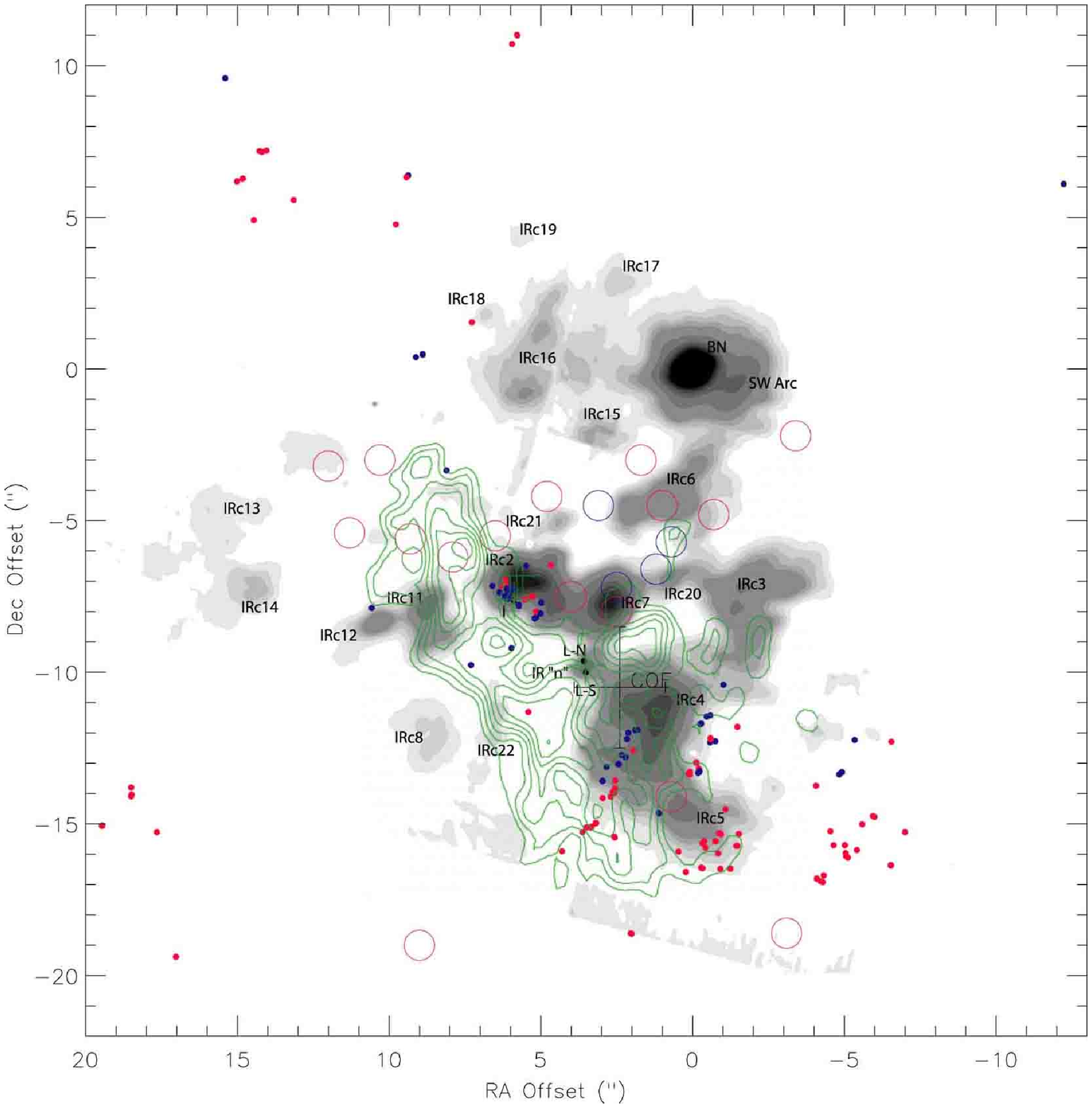}
\caption{Filled contour map of the 12.5~\micron\ mosaic (grayscale, same contour levels as Fig~\ref{fig:bnkl_irsources}) overlaid with \nhthree\ emission at \vlsr  $= 6.79$~\kms\ (green contours) tracing the Orion ``hot core'' \citep{Wilson+00}, \water\ masers \citep[filled dots,][]{Genzel+81,Gaume+98}, and OH maser {\em clusters} \citep[open circles,][]{Johnston+89}.  Red and blue indicate Doppler shift for the maser(s) relative to \vlsr $= 5$~\kms, the approximate velocity centroid of the hot core.  The 18~\kms\ \water\ maser outflow center of expansion (COE) is remarkably close to IR source ``n''.  Notice the significant anti-correlation of \nhthree\ and 12~\micron\ emission, suggesting that most of the mid-IR emission comes from at, or behind the hot core.  The mid-IR emission and maser spots appear to trace bipolar cavities centered on source ``n''. Offsets are relative to BN in arcseconds. }
\label{fig:bnkl_masers_nh3}
\end{figure}

There is a significant anticorrelation between the 12.5
             ~\micron\ emission in our map and \nhthree\ emission from the
              ``hot core'' (Figure~\ref{fig:bnkl_masers_nh3}),
              suggesting that most of the mid-IR sources are {\em at
              the same distance or behind} the hot core.
The very dark regions in our mid-IR mosaic -- the roughly
              triangular region NW of n and the larger region SE of n --
              coincide very nicely with the \nhthree\ emission peaks (the
              correlation is most dramatic in the 6.79~\kms\ channel of
              \nhthree).  
Source n sits precisely at the waist of this
              emission.
The mid-IR and \nhthree\ contours immediately surrounding source n are
              highly suggestive of bipolar cavities, with opening angles of $\sim 90$\arcdeg, aligned nearly
              perpendicular to the long axis of source n.
In addition, most of the OH and \water\ masers in the
              entire BN/KL region can be found within an extension of
              the conical regions defined by these cavities (Figure~\ref{fig:bnkl_masers_nh3})
              suggesting that these regions have been cleared by an
              outflow(s) from source n and that this outflow(s) is responsible for the masers.

The bipolar cavity defined by the mid-IR, \nhthree, and maser
              emission, along with the elongated structure of n at 12
             ~\micron, strongly suggests that n has a circumstellar
              disk with a rotation axis oriented NE--SW.
The unresolved near-IR flux from n would be from the
              stellar photosphere and/or hot inner portion of the disk. 
              The 12~\micron\ emission presumably arises in the outer
              portions of the disk (at hundreds of~AU), where temperatures
              are a few hundred Kelvin, and hence could be
              resolved in our data.  
The FWHM at 12~\micron\ along source n's long axis
              (PA $= 131$\arcdeg) is 0.57\arcsec.  Source n is unresolved
              along the short axis of mid-IR emission.  
If we assume a distance of 450 pc to OMC 1, source n must be
              255~AU in diameter and $< 167$~AU in projected height,
              leading to a disk inclination of $> 50$\arcdeg.  
The disk cannot be edge-on, however, or source n would
              not be detected at near-IR wavelengths owing to extinction
              of the stellar photosphere and inner disk by the outer,
              flared regions.  
The polarization angle and linear fraction for source n at 3.8~\micron\ \citep{Minchin+91,Dougados+93} is consistent with this disk interpretation.
The elongated radio source L, bisected perfectly by the
              mid-IR disk, may be emission from an ultra-compact bi-polar \ion{H}{2} region
              above and below the disk or perhaps a jet and counter-jet.
 The rotation axis of the disk would be along PA $\approx 41$\arcdeg , in near perfect alignment with source L (N and S) and the bipolar cavities to the NE and SW.
Because of its significant radio continuum emission, source n has
              already been identified as a possible contributor to the
              energetics of the BN/KL region \citep{Menten+Reid95}.
Our data, coupled with the \nhthree\ observations of the hot
              core and the maser spots observed throughout BN/KL,
              suggest a disk and outflow morphology, as would be
              expected for a young, high-mass star.

\subsection{Source I and IRc2}

\begin{figure}
\epsscale{0.75}
\plotone{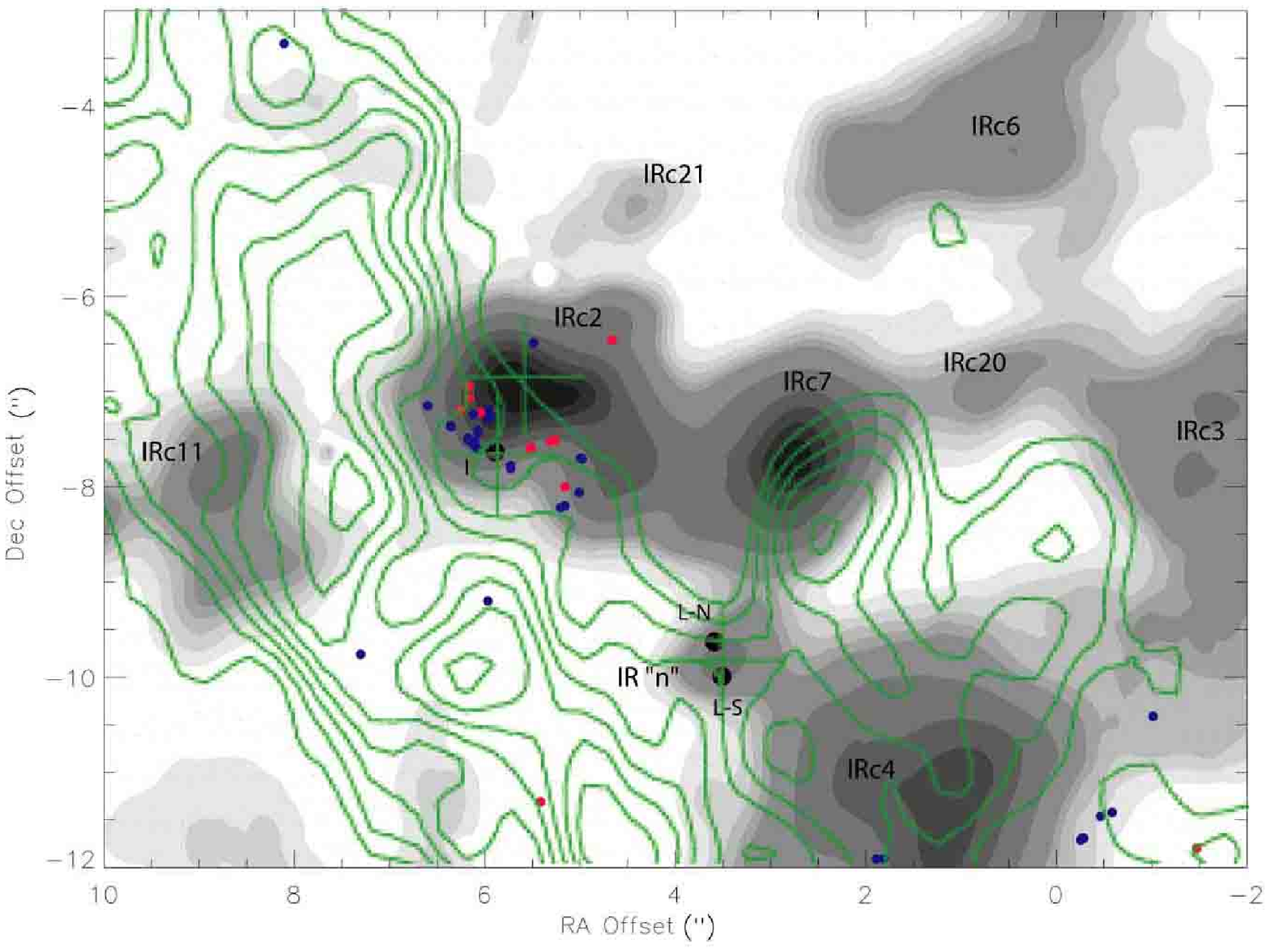}
\caption{Filled contour map of the 12.5\micron\ mosaic (grayscale, same contours as Fig.~\ref{fig:bnkl_irsources}) overlaid with \nhthree\ emission at $v_{lsr} = 5.57$~\kms\ (green contours) tracing the Orion ``hot core'' \citep{Wilson+00}, \water\ masers \citep[filled dots,][]{Genzel+81,Gaume+98}.  Red and blue indicate Doppler shift for the maser(s) relative to $v_{lsr} = 5$~\kms, the approximate velocity centroid of the hot core.  The ammonia emission strongly suggests a cavity surrounding radio source I with a ``thumb'' of dense material occulting the southern part of IRc2.  The ``shell masers'' surrounding source I may arise in shocked regions in the cavity wall.  Offsets are relative to BN in arcseconds.}
\label{fig:IRc2_masers_nh3}
\end{figure}

The strong radio continuum from source I, indicative of
              an ultra-compact \ion{H}{2} region, suggests that it also
              is a significant energy source for the BN/KL region
              \citep{Menten+Reid95}.
Source I has not been detected in the IR, however, indicating
              that if it is indeed a hot, young star, it must be
              obscured by over 60 visual magnitudes of dust, placing it within or
              on the far side of the hot-core \citep{Gezari+98}.  
Observations of the 86 GHz SiO masers led to an interpretaion, at first, of an outflow from source I aligned with PA $\sim -45$\arcdeg\ \citep{Greenhill+98}.  
New VLBA observations, however,  suggest that the SiO masers arise in a
              disk seen nearly edge-on with a rotation axis at PA $\sim
              45$\arcdeg \citep{Greenhill+03}.
In this picture, the water masers centered on source I
              (the ``shell'' masers) are due to a wide angle bipolar
              outflow from I impacting the surrounding dense natal
              material \citep{Greenhill+03}.
Indeed, the shell masers trace very closely the \nhthree\
              contours surrounding source I and the mid-IR contours in
              our mosaic (Figure~\ref{fig:IRc2_masers_nh3}), suggesting
              that they arise in the walls of a cavity
              excavated out of the hot core gas by bipolar outflows from source I    
          \citep{Wilson+00}.

The nature of the IRc2 sources (A -- E) is not entirely
              clear.  
Because of their proximity to source I on the sky and the fact that the
              mid-IR emission comes from a location which is at or behind the hot core (in
              which source I is most likely embedded), it seems that
              the IRc2 sources are associated with source I.
The silicate extinction factor (9.8~\micron\ line-to-continuum ratio) for IRc2 is $\sim 50$ -- higher than any other source in BN/KL \citep{Gezari+98}.  Due to the correlation with source structure, the extincting grains are probably local to IRc2 itself.
The dereddened flux densities in the mid-IR for IRc2 indicate a color temperature of $\approx 240$ K and a total luminosity of $L = 1000 \pm 500$\lsun\ \citep{Gezari+98}.  
At least part of the strong silicate absorption could be due to the hot core, which clearly overlaps some of the mid-IR emission (Figure~\ref{fig:IRc2_masers_nh3}).  

The IRc2 sources also show significant polarization in the near-IR \citep{Minchin+91,Dougados+93}.  The high-resolution \Lp\ polarimetry map by \citet{Dougados+93} clearly shows linear polarizations of $10--15$\% for the IRc2 sources with polarization angles roughly perpendicular to the direction of source I.  This naturally leads to the interpretation that some of near-IR emission from IRc2 is due to scattering of radiation from source I.

The wavelength dependent morphology of IRc2 (Fig.~\ref{fig:IRc2_12-4}) displays two important characteristics:
\begin{enumerate}
\item the resolved structure of IRc2 at 12.5~\micron\ becomes more unresolved at shorter wavelengths, culminating in the point-like nature of sources A -- D  at \Lp; and
\item the 3.8 -- 5.0~\micron\ sources A -- D generally appear at the periphery of the 12.5~\micron\ emission.
\end{enumerate}
How to interpret these characteristics, however, is not at all clear:  the first characteristic suggests that these sources are embedded protostars (as indicated by the mid-IR color of IRc2); while the second is indicative of an externally heated clump of gas and dust (as suggested by the near-IR polarimetry results).  In addition, IRc2 appears to be partially extincted by a tongue of dense foreground material which is revealed by a small bay in the extended 12~\micron\ emission at $(\Delta \alpha, \Delta \delta) = (5.25,-7.80)$, and coincides with a small protrusion in the \nhthree\ emission (Figure~\ref{fig:IRc2_masers_nh3}), so patchy extinction plays a strong role in the observed morphology as well.  

If sources A -- E were indeed embedded protostars, then the total IR luminosity for IRc2, $L = 1000 \pm 500$~\lsun\,  indicates that their masses would likely be in the 3 -- 8~\msun\ range.  This would imply that, in the 500-AU-diameter region where the IRc2 cluster sources are located, the volume averaged density would be larger than $10^{10}$ \cmthree\.  This would represent a compression by at least a factor of $\sim$50 in all three dimensions relative to the density of a typical star-forming cloud core, $10^{5}$ \cmthree\, and the original cloud core would have to have been larger than the whole BN/KL region.  Therefore, while it is probably not impossible to get a group of massive stars forming so close to each other (the challenge is avoiding the centrifugal barrier during the collapse), it seems far more likely that sources A -- E are patches of reflecting dust rather than embedded protostars.

In addition, the near-IR polarization and wavelength-dependent morphology of IRc2 from 3.8 to 12.5~\micron\ point strongly to external heating and illumination. Indeed, it seems very likely  that IRc2 is illuminated and heated by radiation and outflow-driven shocks from source I---especially if the outflow is oriented along a NW--SE axis \citep{Greenhill+98}.  Assuming that the luminosity of source I is a reasonable $L \sim 10^4$~\lsun, corresponding to a 10 -- 20~\msun\ star, IRc2 would have to intercept $\sim 10$\% of its total energy to account for its luminosity.  Given the
apparent relative placement of the IRc2 components and source I, this appears  to be a plausible number.
It is possible, however, that both external and internal heating play a role in the emission from IRc2, as would be expected if a low-mass, dense protostellar cocoon were placed very near an energetic source.  Such objects would be the evolutionary forebears of the ``proplyds'' seen near the Trapezium.

It is interesting to note that IRc2 A is clearly coincident with the shell \water\ masers  NE of source I, especially at 5.0~\micron\ (Figure~\ref{fig:IRc2_masers_nh3}).  If the outflow from source I is instead oriented along a NE--SW axis, as suggested by \citet{Greenhill+03}, then IRc2 A may be thermal emission from the cavity being excavated out of the hot core by source I. The rest of the IRc2 sources might be shadowed by I's circumstellar disk and hence would be self-luminous and/or heated by source n and BN.

\subsection{The OMC 1 Outflows}

\begin{figure}
\epsscale{0.75}
\plotone{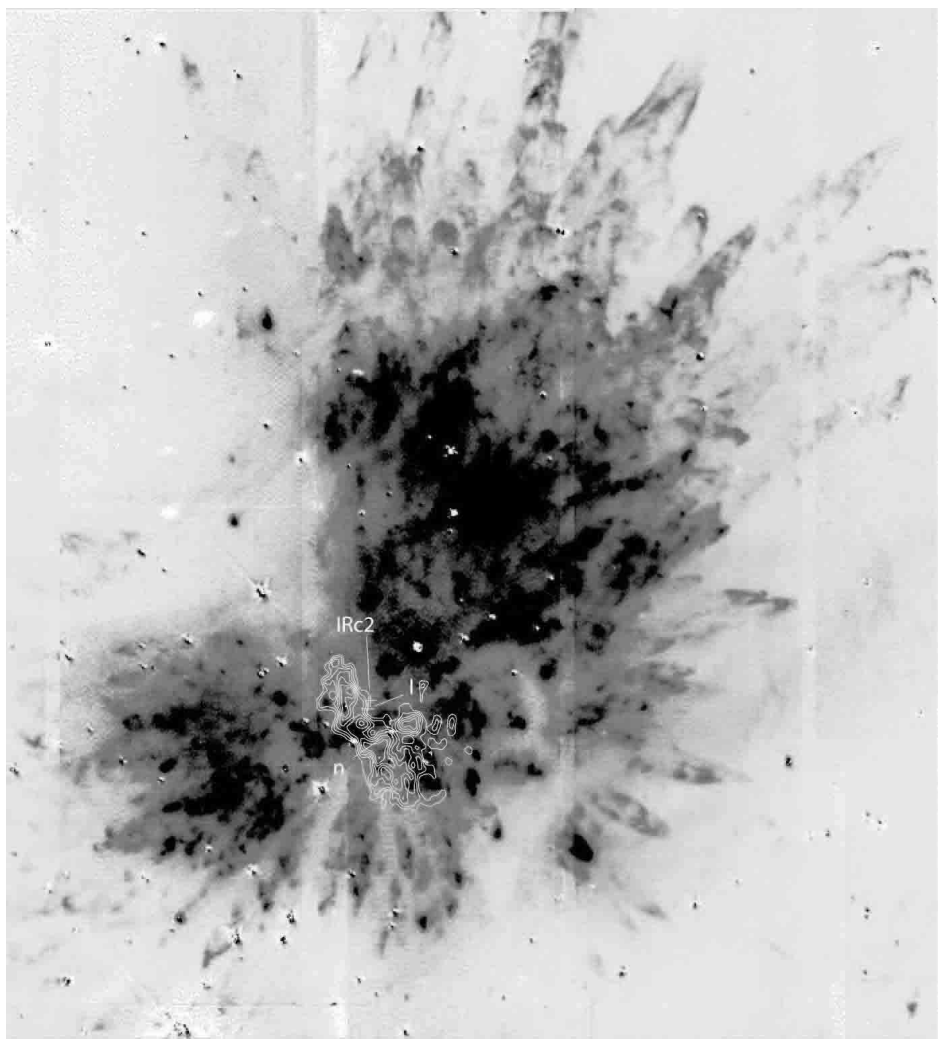}
\caption{\htwo\ image of the OMC 1 region (Subaru Observatory)
overlaid with \nhthree\ emission contours tracing the Orion hot core \citep{Wilson+00}.  The image is roughly 2\arcmin\ $\times$ 2\arcmin\ in size.}
\label{fig:OMC 1_H2_NH3}
\end{figure}

As discussed in Section~\ref{sect:Intro}, observational evidence from a variety of tracers indicates that there are (at least) two outflows emerging from OMC 1.
Radio CO and near-IR \htwo\ line emission show a fast, poorly
              collimated bipolar outflow
              along a NW--SE axis.  
On arcminute scales, the \htwo\ fingers appear to be
              erupting roughly orthogonal to the
                \nhthree\ ridge (Figure~\ref{fig:OMC 1_H2_NH3}).
Closer in,  the shocked regions are correlated, but not
              coincident, with clouds observed at 12~\micron\ (Figure~\ref{fig:BNKL_12-N215-H2_ann}), suggesting that
              the \htwo\ emission arises from the shocked surfaces of
              warm clouds.
A plausible explanation is that the fingers represent
                high velocity ejecta moving relatively unimpeded into
                low density regions NW and SE of the core, whereas the
              masers represent
               strong shocks at the locations where the ejecta encounter some dense gas in the
              OMC 1 ridge.

Recent VLBA observations of the SiO masers have been interpreted
as evidence that the disk around source I has a minor axis oriented
towards PA = 45\arcdeg\ \citep{Greenhill+03},  roughly 90\arcdeg\ from the previous 
interpretation by \citet{Greenhill+98}.  If this is correct, then
there is no obvious source for the high-velocity ``explosive'' outflow 
originating from the hot core and aligned NW--SE as traced by CO and
H$_2$ emission.  But there are several alternative possibilities:
First, it is possible that the disk orientation has flipped by about 
90\arcdeg\ within the last 1,000 years since the production of the
H$_2$ fingers.  Obviously, this hypothesis invokes a set of rather peculiar circumstances.  Such an orientation change may be induced by the close
passage of another massive star, or the sudden accretion of 
material with radically different angular momentum orientation.  
Second, perhaps one (or both) of the sources, I and n, suffered a major
eruption that was more or less isotropic.  As this flow ran into the
NE--SW oriented ridge of dense gas traced by NH$_3$ observations, 
it was blocked. Some of the ejecta may have been deflected along the density
gradient to produce a SE--NW flow.  The problem with this interpretation 
is that it does not explain why the source I disk did not cast a mechanical
shadow along its plane.  Finally, it is possible that the recent 7 mm
observations of source I do not trace dust emission from a disk.  
Rather, this emission traces a dense thermal jet oriented SE--NW, exactly
along the major axis of the high velocity CO outflow and H$_2$ fingers.
In this latter picture, the distribution of SiO masers within an arc second
of source I do not trace the surface of a disk, but rather the shear layer
formed where the jet interacts with a thick torus or cored-sphere
of circumstellar material.  While the geometry of this hypothesis is compelling, the observed velocities of the SiO masers and the CO and \htwo\ fingers do not match:  the SiO masers are geneerally red-shifted to the NW of source I and blue-shifted to the SE while the CO and \htwo\ emission shows exactly the opposite velocity distribution.

We suggest the following scenario for the BN/KL region.  
  Most of the H$_2$O and OH masers (but not the ``shell masers'' 
  in the immediate vicinity of source I) are produced by 
  a relatively steady bipolar outflow from source n \citep[corresponding to the
  ``18 km s$^{-1}$'' flow,][]{Genzel+81} which has cleared two cavities to the NE and SW.
  The H$_2$O maser motions suggest that this outflow has been active for
  at least 3000 years.  The relationship between the explosive outflow
  associated with the H$_2$ fingers and oriented NW-SE and the outflow
  from source n is not clear, though both have probably contributed to
  the production of high velocity gas in the BN/KL region.  These flows
  have similar dynamical ages and it is not clear which flow stared first.
  If the explosive ejecta moved relatively unimpeded through the medium, 
  then the flow from source n may have cleared a cavity before the  
  the explosive event traced by the H$_2$ fingers. The limited
  spatial extent of the ``shell masers'' suggests that source I may be younger 
  than source n.

An interesting feature of the NE and SW outflow cavities
              is the apparent anti-correlation of OH and \water\ masers
              (Figure~\ref{fig:bnkl_masers_nh3}):  the NE cavity is
              dominated by OH masers and the SW by \water\ masers.  
This discrepancy might be because of the photodissociation of
              \water\ in the NE cavity by strong UV radiation from
              source I, n, BN, or perhaps even \thetaonec.  
The SW cavity may be shielded from any UV flux from
              source I and BN (though not n) allowing \water\ to
              survive and form maser spots.

\subsection{IR source ``t''}

The mid-IR and H$_2$ emission north of IR source ``t'' (SE of IRc2; 
Figure~\ref{fig:BNKL_12-N215-H2_ann}) trace a semi-circle of H$_2$ emission with a diameter
of $\sim$ 3\arcsec\ (1370~AU at 450 pc). The 12.5 $\mu$m emission
south of IRc11 forms a foot-shaped ridge following the
H$_2$ emission north of source t.
In addition, the mid-IR sources IRc8 and 22 appear to continue the circular
              structure  around source t to the south.


\htwo\ emission can arise in both shock-excited regions
              and gas heated to a few thousand K by UV radiation.
The circular structure around source t is highly
              suggestive of a PDR expanding into a higher density
              cloud -- which can be accommodated if source t is a hot star on the near
              side, and close to the surface, of the hot core.
If this hypothesis is correct, then source t must also be a
              significant source of UV flux, though not necessarily in
              the Lyman continuum since \htwo\ is excited by soft-UV
              photons.  Thus, source t may be a late B or early A star.
As the PDR advances outward from source t, the radiation front illuminates over-dense
              regions which then emit in the mid-IR.  
We postulate that the local extinction is somewhat higher
              to the south of t, blocking 2~\micron\ \htwo\ emission from the
              southern part of the PDR, while allowing 12.5~\micron\
              emission from IRc8 and  22 to pass.
The interface of the PDR with the IRc11 ``foot'' seems to
              be viewed nearly edge-on.

\section{Summary \& Conclusions}
\label{sect:Summary+Conclusions}

In this paper we have presented the highest spatial
          resolution map at 12.5~\micron\ of the BN/KL region to date.
Despite relatively poor flux calibration owing to variable
          atmospheric transmission we were able to detect 6 new
          sources and resolve new morphologies for existing sources leading to the following results:
\begin{itemize}
\item
IRc2 is resolved into multiple sources, as observed at $\sim 4$
         ~\micron \citep{Dougados+93}.  
The point-like nature of sources A -- D at 3.8~\micron\ coupled with the relatively high color temperature of IRc2 in the mid-IR, seems to suggest that these are self-luminous sources.  Near-IR polarization maps, however, suggest that IRc2 is externally illuminated.   Comparison of our 12.5~\micron\ map with previous datasets clearly show that the shorter wavelength sources all appear at the periphery of the 12.5~\micron\ emission.  This lends some weight to the picture in which IRc2 is illuminated and heated by  source I, but we cannot rule out the possibility that IRc2 is a small group of nearby protostars.

\item
The elongated structure of near-IR source n at 12~\micron,
          coupled with the geometry of the mid-IR and \nhthree\
          emission, and the location of \water\ and OH maser spots, is
          highly suggestive of a disk with rotation axis along a NE--SW
          axis on the sky.  
Thus, source 
n may be responsible for the low-velocity 18~\kms\ flow emerging 
from OMC 1.  

\item
In addition, we have discovered an arc of emission associated
          with the BN object and some interesting morphology for
          IR source t, SE of IRc2.  The BN SW arc is a curious feature that requires follow-up imaging and spectroscopy at other wavelengths to determine its true nature.

\end{itemize}

Owing to the extreme
          obscuration on the line of sight, our mid-IR observations are only sampling the most massive end
          of the IMF in BN/KL.  
The stellar density in the region is likely to be very high (as
          suggested by the source I--IRc2  group), perhaps even
          greater than the Trapezium cluster.  This high stellar density may be due to environment or evolutionary state (or both):  It is possible that the Trapezium Cluster also began with a much higher stellar density.
Additional deep mid-IR observations, especially at 4 and 8~\micron\ accompanied by 20~\micron\ images to determine spectral energy distributions, are required to determine the full stellar content of this rich environment.

\onecolumn
\small

\acknowledgements

The authors would like to thank an anonymous referee whose comments improved this manuscript substantially.  This research has been supported by a cooperative agreement through the Universities Space Research Association (USRA) under grant \# 85502-02-02 to M. Morris and by NASA Astrobiology grant NCC2-1052 to the University of Colorado.  

Data presented herein were obtained at the W.M. Keck Observatory, which is operated as a scientific partnership among the California Institute of Technology, the University of California and the National Aeronautics and Space Administration. The Observatory was made possible by the generous financial support of the W.M. Keck Foundation.  The authors wish to recognize and acknowledge the very significant cultural role and reverence that the summit of Mauna Kea has always had within the indigenous Hawaiian community.Ê We are most fortunate to have the opportunity to conduct observations from this mountain.

Some of the data presented in this paper was obtained from the Multimission Archive at the Space Telescope Science Institute (MAST, \url{http://archive.stsci.edu/}).  STScI is operated by the Association of Universities for Research in Astronomy, Inc., under NASA contract NAS5-26555.  Support for MAST for non-HST data is provided by the NASA Office of Space Science via grant NAG5-7584 and by other grants and contracts.



\clearpage




\clearpage
\renewcommand{\arraystretch}{0.6}
\begin{deluxetable}{lcccccll}
\tabletypesize{\footnotesize}
\tablewidth{0pt}
\tablecaption{Mid-IR Sources in BN/KL\label{table:sources}}
\tablehead{
\colhead{}  &  \colhead{Offset\tablenotemark{a}}  &  \colhead{Offset\tablenotemark{a}} &
\colhead{Positional} & \colhead{}  & \colhead{Aperture } &  \colhead{}  \\
\colhead{Name}  &  \colhead{RA (\arcsec)}  &  \colhead{Dec. (\arcsec)} &
\colhead{Error (\arcsec)} & \colhead{Flux (Jy)\tablenotemark{b}}  & \colhead{Radius (\arcsec)} &  \colhead{Notes} & \colhead{Refs.}  \\
}
\startdata					
\objectname{BN}	&	0.0	&	0.0	&	0.1	&	[600]	 &	2.025	& & 2,3 \\
\objectname{BN SWarc }		&	-1.4	&	-0.6	&	0.1	&		&  &	Elongated along PA = 131\arcdeg  & 1  \\
\objectname{IR n }		&	3.5	&	-9.8	&	0.1	&	14.2	&	0.81	&  & 3,4 \\
\objectname{IRc2 A+B }		&	5.7	&	-7.0	&	0.1	&	[120]	&	1.215 & 
Single 12~\micron\ source between  A \& B & 5 \\
\objectname{IRc2 C }		&	4.6	&	-7.8	&	0.1	&		&		& & 6  \\
\objectname{IRc2 D }		&	5.0	&	-7.0	&	0.1	&		&		& & 6 \\
\objectname{IRc2 E }		&	4.7	&	-6.3	&	0.1	&		&		& & 1,6 \\
\objectname{IRc3N }		&      -1.6	&      -7.2	&    0.1  &		&		& & 1 \\
\objectname{IRc3S }		&      -1.4	&      -7.9	&    0.1  &		&		& & 1 \\
\objectname{IRc4 }		&	1.1	&	-11.1	&	0.1	&		&		&
Position here is peak at NW vertex. & 3,7  \\
\objectname{IRc5 }		&	0.5	&	-14.5	&	0.2	&	19	&	0.891 &  & 3,7 \\
\objectname{IRc6 }		&	0.4	&	-4.5	&	0.1	&		&		&  Central source & 3,8 \\
\objectname{IRc6N }		&	-0.2	&	-3.2	&	0.1	&		&		& & 3 \\
\objectname{IRc6E }		&	2.0	&	-4.8	&	0.1	&		&		& & 1 \\
\objectname{IRc7 }		&	2.8	&	-7.8	&	0.1	&	108	&	1.215  &  & 3,8 \\
\objectname{IRc8 }		&	8.4	&	-12.2	&	0.2	&		&		&  & 3,8 \\
\objectname{IRc11 }		&	8.8	&	-7.9	&	0.2	&	18.7	&	0.81	&
Position and flux are for  & 3 \\
	&		&		&		&		&		& 
	primary elliptical source \\
\objectname{IRc12 }		&	10.3	&	-8.4	&	0.2	&	10.1	&	0.81  & & 3 \\
\objectname{IRc13 }		&	14.5	&	-4.8	&	0.3	&		&		&
Diffuse, broken structure	& 3 \\
\objectname{IRc14 }		&	14.5	&	-7.2	&	0.3	&		&		&
Diffuse, broken structure	& 3 \\
\objectname{IRc15 }		&	3.1	&	-2.3	&	0.1	&		&		&
Position uncertain (edge)	& 3 \\
\objectname{IRc16N }		&	4.9	&	1.3	&	0.2	&		&		& & 1 \\
\objectname{IRc16S }		&	5.5	&	-0.8	&	0.2	&		&		& & 1 \\
\objectname{IRc16W }		&	4.0	&	0.2	&	0.1	&		&		& & 1 \\
\objectname{IRc17 }		&	2.5	&	2.8	&	0.1	&		&		&  & 3 \\
\objectname{IRc18 }		&	6.8	&	1.8	&	0.2	&	2.6	&	0.729  &  & 3 \\
\objectname{IRc19 }		&	5.7	&	4.2	&	0.2	&		&		&
Associated with NIR "q" ? &1,4 \\
\objectname{IRc20 }		&	0.7	&	-6.9	&	0.1	&	7.1	&	0.729  &  & 1\\
\objectname{IRc21 }		&	4.4	&	-5.0	&	0.1	&	5.2	&	0.729  &  & 1\\
\objectname{IRc22}	&	6.4	&	-11.3	&	0.2	&		&	&
Faint N--S ridge  & 1	\\
\enddata
\tablenotetext{a}{Offset from BN  (RA = 05h35m14.117s,  DEC = -05\arcdeg22\arcmin22.90\arcsec, J2000) }
\tablenotetext{b}{Total sky-subtracted flux measured in the aperture listed; the sky value was determined from an annulus just beyond the aperture.  $2\sigma$ error for total flux is 10\% -- 27\%, see text.  Fluxes in brackets (BN and IRc2) are from \citet{Gezari+98} and were used for calibration.}
\tablerefs{(1) This work; 
(2) \citet{Becklin+Neugebauer67}; 
(3) \citet{Gezari+98}; 
(4) \citet{Lonsdale+82};
(5) \citet{Chelli+84};
(6) \citet{Dougados+93};
(7) \citet{Rieke+73};
(8) \citet{Downes+81}
}
\end{deluxetable}

\end{document}